\documentclass{article}
\title{{\bf Quantum Physics in a different ontology}\\
\begin{tabular}{c}
\normalsize  Nalin de Silva
\\
\vspace*{-1.1cm}\\
\normalsize   Department of Mathematics, University of Kelaniya, Kelaniya, Sri Lanka\\
\vspace*{-1.1cm}\\
\end{tabular}
\date{}
}
\usepackage{fancyheadings}
\usepackage{amsmath,amssymb,graphicx}
\begin{document}
\maketitle
\section*{Abstract}
It is shown that neither the wave picture nor the ordinary particle picture offers a satisfactory explanation of the double-slit experiment. The Physicists who have been successful in formulating theories in the Newtonian Paradigm with its corresponding ontology find it difficult to interpret Quantum Physics which deals with particles that are not sensory perceptible. A different interpretation of Quantum Physics based in a different ontology is presented in what follows. According to the new interpretation Quantum particles have different properties from those of Classical Newtonian particles. The interference patterns are explained in terms of particles each of which passes through both slits.     

\section{INTRODUCTION}

Planck introduced his ideas on quanta or packets of energy towards the end of the nineteenth century. In that sense Quantum Physics is more than one hundred years old. From the very beginning Quantum Physics came up with strange phenomena that made the Physicists to disbelieve what they themselves were proposing to understand the new features that were being observed.  
\vspace*{5mm}

The so-called double-slit experiment$^1$  continues to baffle the Physicists who are glued to twofold two valued logic that is behind the Newtonian paradigm. As it was one of the most fundamental experiments that they could not understand in Quantum Physics the Nobel Prize winning Physicist Richard Feynmann once declared that no body understood Quantum Physics! This statement by Feynmann makes one to delve into the meaning of understanding. In other words one has to understand what is meant by understanding. However, it is clear that if one is confined to an ontology based in twofold formal logic, and linear thinking one would be confused by a statement such as understanding what is meant by understanding.  A decade ago the intellectuals who were only familiar with linear thinking and not with cyclic thinking would have left deliberations into such statements to whom they call mystics, as such statements did not come within the ``rational'' way of thinking. However, in this paper we would not attempt to understand what is meant by understanding.
\vspace*{5mm}

The principle of superposition which was familiar to Classical Physicists as well, has taken an entirely different meaning with respect to Quantum Physics. The essence of the principle can be explained as follows. If $x$ and $y$ are two solutions of what is called a linear differential equation then $x+y$ is also a solution of the same differential equation. This is a simpler version of what is generally known as the principle of superposition. In Classical Physics two magnets giving rise to two different magnetic fields would combine to give one magnetic field, and a compass that is brought to the resulting magnetic field would respond to the resulting field, and not to the field of any one of the magnets. It has to be emphasised that a magnet is in only one state, corresponding to the respective magnetic field and it is the two fields of the two magnets that combine to give one field though one would not find a single magnet that gives rise to the resultant field. We could describe this phenomenon as that of two or more becoming one. However, in the Quantum world things are different, and the principle of superposition has an unusual interpretation.

\section{THE WAVE NATURE OF PARTICLES}
In order to discuss the new interpretation of the principle of superposition we first consider the so called double-slit experiment where a stream of electrons (in general, particles or photons) is made to pass through two slits and then to strike a screen. If both slits are open an interference pattern is observed on the screen. Now in Quantum Physics it is said that particles such as electrons posses wave properties and photons (light) exhibit particle properties in addition to their respective ``normal'' properties. Interference patterns are supposed to result from wave properties and according to the Physicists the wave theory successfully explains the formation of such patterns in the case of a stream of particles fired from a source to strike the screen after passing through the slits. The Physicists would claim that the double-slit experiment demonstrates that particles such as electrons do exhibit wave properties.
\vspace*{5mm}

The double-slit experiment has been carried out with only one electron passing through the slits one at a time$^2$  (electrons at very low intensities) instead of a stream of particles released almost simultaneously to pass through the two slits. Even at very low intensities interference patterns have been observed after sufficiently large number of electrons had been fired from the source. The Physicists have been puzzled by this phenomenon. In the case of several electrons passing through the slits simultaneously it could be explained using the wave properties of the particles, in other words resorting to the wave picture. Unfortunately in the case of electrons being shot one at a time this explanation was not possible as what was observed on the screen was not a faint interference pattern corresponding to one electron but an electron striking the screen at a single point on the screen. These points in sufficiently large numbers, corresponding to a large number of electrons, finally gave rise to an interference pattern. The wave nature is only a way of speaking, as even in the case of large number of particles what is observed is a collection of points and not waves interfering with each other.  
\vspace*{5mm}

The Physicists also believe that an electron as a particle could pass through only one of the slits and a related question that has been asked is whether it was possible to find out the slit through which an electron passes on its way to the screen. Various mechanisms, including ``capturing'' the electron using Geiger counters, have been tried to ``detect the path'' of the electron, and it has been found that if the particular slit through which the electron passed was detected then the interference patterns were washed out. In other words determining the particle properties of the electron erased its wave properties. Bohr, who was instrumental in formulating the Copenhagen interpretation$^3$, was of the view that one could observe either the particle properties or the wave properties but not both, and the inability to observe both wave and particle properties simultaneously came to be referred to as complementarity. The experiments that attempted to determine the slit through which the electron passed were known as which-way (welcherweg) experiments as they attempted to find the way or the path of the particle from the source to the screen. The outcome of these experiments made it clear that the which-way experiments washed out the interference patterns. It was believed that at any given time the electrons exhibited either the particle properties or wave properties but not both. 
\vspace*{5mm}

However, what the Physicists failed to recognize was that in the case of one electron shot at a time there was no weak interference pattern observed on the screen for each electron thus illustrating that a single electron did not exhibit any wave properties. The electron strikes the screen at one point, and it is the collection of a large number of such points or images on the screen that gave the interference pattern. In the case of a stream of electrons fired to strike the screen each electron would have met the screen at one point and the collection of such points or images would have given rise to an ``interference pattern''. Thus we could say that the interference patterns are obtained not as a result of the ``wave nature'' of electrons but due to the collectiveness of a large number of electrons that strike the screen. The ``wave nature'' arises out of ``particle'' properties and not due to ``wave properties''. Afshar$^4$  comes closer to this view when he states ``in other words, evidence for coherent wave-like behavior is not a single particle property, but an ensemble or multi-particle property''. We are of the opinion that in the double-slit experiments no wave properties are observed contrary to what is generally believed. It is the particle properties that are observed, though not necessarily those of ordinary classical particles.     
\vspace*{5mm}

As a case in point this does not mean that a particle in Quantum Physics has a definite path from the source to the screen through one of the slits, as could be expected in the case of classical particles. For a particle to have a path it should posses both position and momentum simultaneously. A path at any point (assuming that it is a continuous path without cusps and such other points) should have a well defined tangent. In the case of a particle moving, the direction of the velocity (and the momentum) of the particle at any given point defines the unit tangent vector to its path. Conversely the tangent to the path at any point defines the direction of the velocity and the momentum of the particle at that point. However, according to the Uncertainty Principle, both the momentum and the position of a particle cannot be determined simultaneously, and if the position is known then the momentum cannot be determined. Without the momentum the direction of the velocity of the particle and hence the tangent vector cannot be known implying that a continuous curve is not traced by a particle in space. On the other hand if the momentum of the particle is known then only the direction and magnitude of the velocity (momentum) and properties of other non conjugate observables such as spin of the particle are known, without the position being known. Thus the particle can be everywhere, with variable probabilities of finding the particle at different points, but at each point the particle being moving in parallel directions with the same speed. However, as will be explained later, this does not mean that we could observe the particle everywhere.       
\vspace*{5mm}

In the light of the uncertainty principle it is futile to design experiments to find out the path of a particle. The so-called which-way experiments have been designed to detect the slit through which the particle moves, on the assumption that the particle moves through one slit only. However, in effect there is no path that the particle follows and it is not correct to say that the particle passes through one of the slits. The which-way experiment actually stops the particle from reaching the screen and hence there is no possibility of obtaining any ``interference pattern''. It is not a case of observing particle properties destroying the wave properties of matter, but an instance of creating a situation where the particle is either not allowed to strike the screen or to pass through only one slit deliberately.  In effect it is the particle properties exhibited at the screen that are cut off.  
\vspace*{5mm}

What is important is to note that interference patterns are observed only if both slits are kept open, and also if the particles are free to reach the screen. If one slit is closed or obstacles are set up in the guise of which-way experiments or otherwise, so as not to allow the particles to reach the screen then no interference patterns are observed. The most important factor is the opening of the two slits. In the case of which-way experiments as well, what is effectively done is to close one of the slits as particles through that slit are not allowed to reach the screen. With only one slit open while the other slit is effectively closed with the which-way experiment apparatus, no interference patterns are observed.  
\vspace*{5mm}

The Physicists are obsessed with the idea that a particle can be only at one position at a given time, backed by the ontology of day to day experience. While this may be the experience with our sensory perceptible particles (objects) or what we may call ordinary Newtonian classical objects such as billiard balls, it need not be the case with Quantum particles. However, from the beginning of Quantum Physics, it appears that the Physicists have been of the view that a particle can be at one position at a given time whether it is being observed or not. Hence they seem to have assumed that on its ``journey to the screen from the source'' a particle could pass through only one of the slits. They have worked on the assumption that even if both slits are open the particle passes through only one of the slits but behaves differently to create interference patterns as if the particle is ``aware'' that both slits are open. According to the view of the Physicists if only one slit is open the particles having ``known'' that the other slit is closed pass through the open slit and ``decide'' not to form any interference patterns. It is clear that the explanation given by the Physicists for the formation of interference patterns on the basis of the particle picture is not satisfactory. We saw earlier that the explanation given in the wave picture is also not satisfactory as a single electron fired from the source does not form a faint interference pattern on the screen. If the particles behave like waves then even a single particle should behave like a wave and produce a faint interference pattern, having interfered with itself. What is emphasised here is that the final interference pattern is not the sum of faint interference patterns due to single particles, but an apparent pattern formed by a collection of images on the screen due to the particles. There is no interference pattern as such but only a collection of the points where the particles strike the screen, or of the images formed by the particles that were able to reach the screen. The images finally depend on the probability that a particle would be at a given position.   
\vspace*{5mm}

Before we proceed further a clarification has to be made on ``seeing'' a particle at a given position at a given time in respect of the double-slit experiment. In this experiment we are concerned with particles released from a source with a given momentum and given energy. As such according to the uncertainty principle, nothing can be said definitely on the position of these particles, immediately after they leave the source. It can only be said that there is a certain probability that the particle would be found in a certain position. Thus the particle is ``everywhere'' ``until'' it is ``caught'' at some position such as a slit or a screen. Though we have used the word ``until'', time is not defined as far as the particle is concerned as it has a definite energy. It can only be said that there is a certain probability that the particle could be ``seen'' at a given place at a given time, with respect to the observer. The particle is not only everywhere but also at ``every instant''. Thus it is meaningless to say the particle is at a given slit at a given time as neither time nor position is defined for the particle with respect to itself. The particle would meet the screen at some position on the screen at some time but ``before'' that it was everywhere and at every instant. A photon that is supposed to ``move along a straight line'' should not be considered as such, but being at all points along the straight line at ``all times'' ``before'' it interacts with a screen or another particle.
\vspace*{5mm}

The probability of an electron striking the screen at a given point with only one slit open is not the same as that when both slits are open. Thus when a large number of particles strike the screen, the different probabilities give rise to different ``patterns'' which are essentially collection of points where the particles meet the screen. The ``interference patterns'' observed when both slits are open are replaced by ``other patterns'' when one of the slits is closed. The ``interference patterns'' as well as the ``other patterns'' are the results of particle properties, the difference being due to the number of slits that are open. If both slits are closed there is no pattern at all as no particle would reach the screen under such conditions. When one of the slits is open there is a probability that the particle can be at the position where the slit is whereas when both slits are open there is a probability that the particle could be at both the slits ``before'' reaching the screen. When both slits are open, the particle is at both slits and the position is not known while the momentum of the particle is not changed and has the original value with which it was shot. However, when one of the slits is blocked the particle is at the other slit implying that the momentum is not known. These uncertainties of the momentum would carry different particles to different places on the screen, while in the case when both slits are open it is the uncertainties of position that make the particle to strike the screen at different positions.  The difference between the ``interference patterns'' and the ``other patterns'' is due to this.

\section{EXPERIMENTS OF AFSHAR}
Afshar$^5$  has claimed that he was able to demonstrate that an electron or a photon would exhibit both particle and wave properties (Figure 1). He allowed light to pass through two slits and to interact with a wire grid placed so that the nodes were at the positions of zero probability of observing a photon. The photons were not affected by the wire grid as the nodes were at the positions of zero probability and at those positions there were no photons to interact with the grid. The photons were then intercepted by a lens system that was able to identify the slit through which any single photon had passed. According to Afshar the nodes of the grid at the positions of zero probability indicated that the wave properties of the photons were observable while the lens system in detecting the slit through which the photon had passed demonstrated the particle properties of the photons.
\begin{figure}[h]
\begin{center}
\includegraphics[scale=0.75]{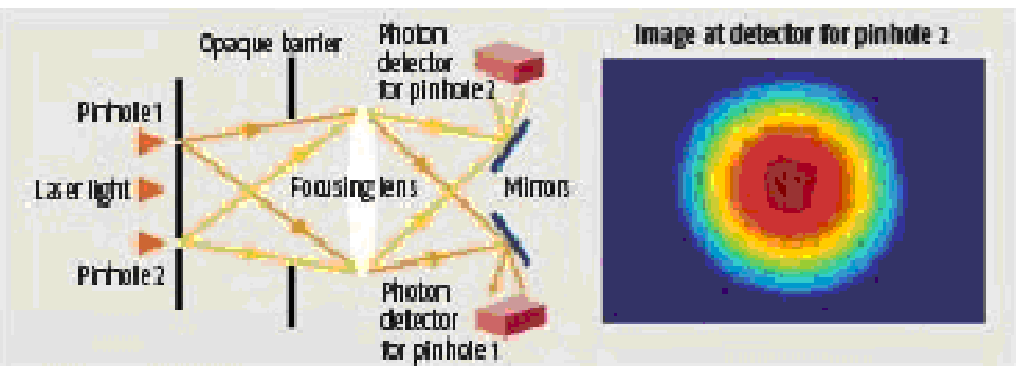}
\includegraphics[scale=0.75]{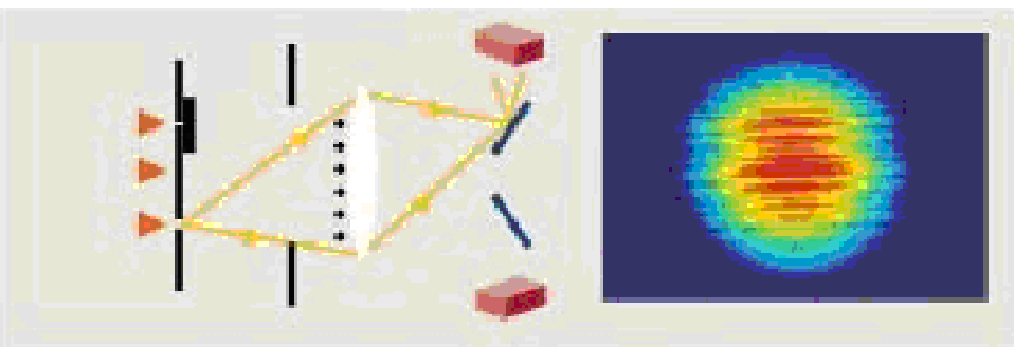}
\includegraphics[scale=0.75]{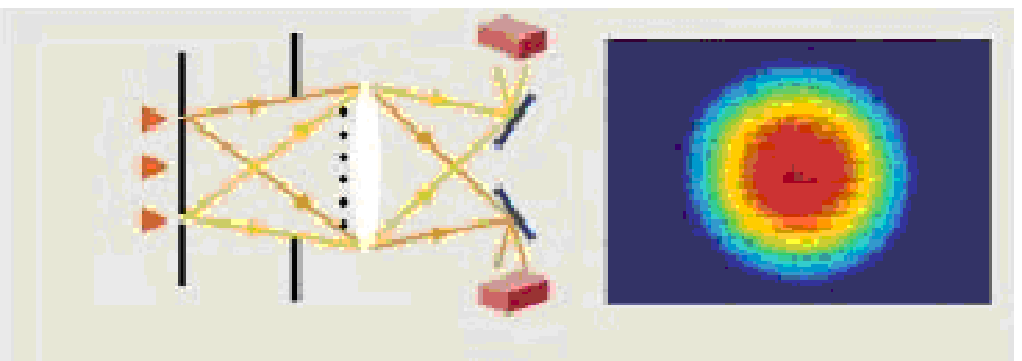}
\end{center}
\caption{{\small The wire grid and the lens system of Afshra, and the corresponding images observed. No interference patterns after the lenses and Afshra claims that the wire grid demonstrates the wave property while the images correspond to the particle property. (Courtesy Afshra)}}
\end{figure}
However, in this experiment, assuming that the lens system detects the slit through which the photon passed, what is observed is again the particle properties of the photons. The wire grid with the nodes at the position of zero probabilities does not interact with the photons, as there are no photons at positions of zero probability to interact with the grid. No so called waves are observed, as there is no screen for the particles to strike. Thus the wire grid has no effect in this experiment and with or without such a grid the lens system would behave the same way.
\vspace*{5mm}

Let us consider what would happen if the wire grid is shifted forwards towards the source, backwards towards the lens system or laterally. As the nodes of the wire grid would be shifted from the positions of zero probability some photons would strike the grid and they would not proceed towards the lens system. Thus the number of photons that reach the lens system would be reduced and there would be a decrease in intensity of light received at the lens. Though Afshar claims that wave properties are observed just by placing a wire grid so that its nodes are at the positions of zero probability, it is not so. 
\vspace*{5mm}

The so called wave properties could be observed only by placing a screen in between the wire grid and the lens system. As we have mentioned above, even then what is observed is a collection of images at the points where the photons strike the screen, and not wave properties as such. In this case as all the photons would have been absorbed by the screen, the lens system would not be able to detect any photons nor the ``slit through which the photons passed''. On the other hand if the screen is kept beyond the lens system then there would not be any photons to strike the screen and hence no ``wave properties''.

\section{EXPERIMENTS AT KELANIYA}
We at the University of Kelaniya have given thought to this problem, and one of my students Suraj Chandana has carried out a number of experiments, which may be identified as extensions of the experiment of Afshar. Chandana and de Silva$^6$  had predicted that if we were to have a single slit and then a screen, instead of the wire grid and the lens system, ``after'' the photons have passed through the two slits, then the photons would pass through the single slit with the same probability as that of finding a photon at the point where the slit was kept. This implied that if the slit was kept at a point where the probability of finding the photon is zero, the photon would not pass through the slit to strike the screen, but on the other hand, if the slit was kept at any other point there was a non zero probability that the photon would pass through the slit, and striking the screen. Thus if a stream of photons is passed through two slits, and ``then'' a single slit, ``before'' striking the screen, depending on the position of the single slit the intensity with which the photons strike the screen would change. Further it implies that these intensities should correspond to the intensities observed in connection with the ``interference patterns'' observed in the case of the standard double-slit experiment, if the positions of the slit were varied along a line parallel (by moving the single slit along a line parallel to the double-slit and the screen)  to the double-slits and the screen. Chandana has been successful in obtaining the results as predicted.  
\begin{figure}[h]
\includegraphics[scale=0.75]{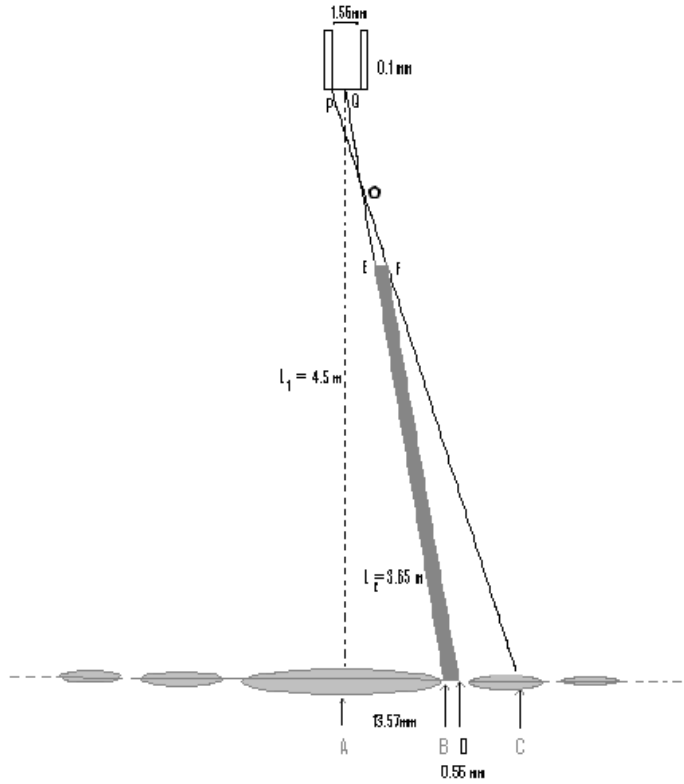}
\caption{{\small The figure represents the aluminium sheet joining the positions of zero probability from a position closer to the double slit to the screen.}}
\end{figure}
In another experiment Chandana$^7$  had an Aluminium sheet of very small thickness joining the points or positions where the probability of finding a photon is zero (positions of zero probability), stretching from the double-slits to the screen as illustrated in the figure 2. As an obstacle placed at a position of zero probability would not affect the photon the Aluminium sheet had no effect on the visible interference patterns on the screen. This experiment was carried out by Chandana with number of Aluminium sheets placed along lines joining the positions of zero probability stretching from the double-slits to the screen. We were not surprised to find that the Aluminium sheets did not interfere with the interference patterns. However, even if one of the sheets is slightly displaced the interference pattern is destroyed as the photons now interact with the sheets at points where the probability of finding a photon is not zero.  
\vspace*{5mm}

These observations are not consistent with the wave picture as a wave would not be able to penetrate the Aluminium sheets without being affected. Even the pilot waves of Bohm  are not known to go through a material medium undisturbed. As we have argued a single electron emitted from the source would not exhibit a faint interference pattern on the screen but a spot or an image having passed beyond the slits. The Physicists are interested in the wave picture to explain the interference patterns as they find it difficult to believe that a particle would pass through both slits simultaneously. Thus they mention of particle properties when they are interested in ``capturing'' particles and of wave properties in explaining phenomena such as the interference pattern.

\section{PRINCIPLE OF SUPERPOSITION IN QUANTUM PHYSICS}
We consider the Quantum entities to be particles though of a nature different from that of Classical Newtonian particles. We have no inhibition in believing that the Quantum particles unlike the Newtonian particles could pass through both slits at the ``same time'', as the logic of different cultures permits us to do so. Physics and in general Mathematics and sciences are based on Aristotelian two valued twofold logic according to which a proposition and its negation cannot be true at the same time. Thus if a particle is at the slit $A$, the proposition that the particle is at $A$ is true and its negation that the particle is not at $A$ is not true, and {\it vice versa}. Therefore if the particle is at $A$ then it cannot be anywhere else as well, and hence cannot be at $B$. This is based on what may be called the Aristotelian- Newtonian - Einsteinian ontology where a particle can occupy only one position at a given time in any frame of reference of an observer.  However, in fourfold logic ({\it catuskoti}) a proposition and its negation can be both true, and hence in that logic it is not a contradiction to say that a particle is at the slit $A$ and at somewhere else (say at the slit $B$) at the ``same instant'' or ``every instant'' Thus according to {\it catuskoti} the particle can be at many places at the same time or at many instants with respect to the observer.       
\vspace*{5mm}

In the case of the double-slit experiment, the momentum of a particle is known, as the particles are fired with known energy, and hence the position is not known. In such a situation Heisenberg's uncertainty principle demands that the position of the particle is not known. The position of the particle is relieved only after a measurement is made to determine the position. Before the measurement, the particle is in a superposition of states corresponding to the positions in space the particle could be found. After the measurement the particle would be found in a definite position (state), collapsing from the superposition of a number of states to that of the definite state.  Before the measurement what could have been said was that there was a certain probability of finding the particle at a given position. Though the particle is in a superposition of states before a measurement is made to find the position, it is in a definite state with respect to the momentum.  
\vspace*{5mm}

In Quantum Mechanics unlike in Classical Mechanics, a state of a system, a particle or an object is represented by a vector in a Mathematical space known as the Hilbert space. The observables such as position, momentum, and spin are represented by what are known as Hermitian operators. If a system is in a state represented by an eigenstate $|\Phi>$ of a Hermitian operator $A$, belonging to the eigenvalue $a$, then the system has the value $a$ corresponding to the observable represented by the Hermitian operator $A$. This is expressed mathematically by  $A|\Phi> = a|\Phi>$.  If $B$ is the conjugate operator of $A$, then the value corresponding to the observable represented by $B$ is not known. All that can be said, according to the standard Copenhagen interpretation, is that if the value corresponding to the observable represented by $B$ is measured, then there is a certain probability of obtaining an eigenvalue of $B$ as the measurement. Before the measurement is made nothing could be said of the value. In plain language this means that if the value of a certain observable is known then the value of the conjugate observable is not known. 
\vspace*{5mm}

However, the state $|\Phi>$ can be expressed as a linear combination of the eigenstates $|\Psi>$ of $B$ in the form $|\Phi>=\sum|c_i\Psi_i>$  where $c_i\in C$, the field of complex numbers. In other words the coefficients of $|\Psi>$'s in the expansion of $|\Phi>$ are complex numbers. The Copenhagen interpretation tells us that when the observable corresponding to $B$ is measured it would result in a state corresponding to one of the $|\Psi>$'s with the measurement yielding the eigenvalue $b$ to which the particular $|\Psi>$ belongs, the probability of obtaining the value $b$ being given by the value of the relevant $|c|^2$. Before the measurement is made nothing can be said regarding the observable corresponding to $B$. According to Bohr, it is meaningless to talk of the state of the system with respect to $B$ as nothing could be observed. There is no knowledge regarding the observable corresponding to $B$ as it has not been observed. The value or the knowledge of the observable is ``created'' by the observer who sets up an experiment to measure the value in respect of $B$. The observed depends on the observer and it makes no sense to talk of an observable unless it has been observed. This interpretation is rooted in positivism as opposed to realism in which the entire corpus of knowledge in Newtonian - Einsteinian Physics is based. This body of knowledge is also based in Aristotelian - Newtonian - Einsteinian ontology.  
\vspace*{5mm}

As a particular case one could refer to the conjugate Hermitian operators in respect of position and momentum of a particle in Quantum Mechanics. When the position of a particle is measured then its momentum is not known. According to the Copenhagen Interpretation, it can only be said that if an apparatus is set up to measure the momentum, the observer would observe one of the possible values for the momentum and that there is a certain probability of observing the particular value. Before the measurement is made the particle has no momentum, as such, and it is meaningless to talk of the momentum of the particle. The observer by his act of observation gives or creates a value for the momentum of the particle, so to speak of. Once the momentum is measured the observer has knowledge of the momentum but not before it. However, after the momentum is measured, the knowledge of the position of the particle is ``washed off'' and hence it becomes meaningless to talk of the position of the particle. The observer could have knowledge only of either the momentum or the position, but not of both. A version of this conclusion is sometimes referred to as the uncertainty principle. 
\vspace*{5mm}

What we have been discussing in the proceeding paragraphs is the principle of superposition.  A particle or a system with its position known is represented by a vector $|\Phi>$in Hilbert space, which is an eigenvector of the Hermitian operator $A$ corresponding to the position. When the position of the particle or the system is known, the momentum is not known. If $B$ is the Hermitian operator corresponding to the momentum, then $|\Phi>$ is not an eigenvector of $B$. However, $|\Phi>$ can be expressed as a linear combination of the eigenvectors $|\Psi>$'s of $B$ though the momentum is not observed. The superposition of the $|\Psi>$'s cannot be observed, and neither can be resolved into observable constituent parts. This is different from the principle of superposition in Classical Physics, where the resultant can be resolved into its constituent parts.  
\vspace*{5mm}

For example as we have mentioned in the introduction the resultant magnetic field due to two magnets can be resolved into its two components and can be observed. One of the magnets can be taken off leaving only one of the constituent magnetic fields. The superposition is there to be observed and if the magnet that was taken off is brought back to its original position the resultant magnetic field reappears. In Quantum Physics the superposition cannot be observed without disturbing the system and when it is disturbed to measure the conjugate variable, only one of the states in the superposition could be observed and we would not have known in advance if that particular state were to appear as a result of the disturbance induced by us.

\section{COPENHAGEN INTERPRETATION}
In Classical Physics, as we have already stated, superposition is there to be observed. However, in Quantum Physics the superposition cannot be observed, and further unlike in Classical Physics interpretations are required to ``translate'' the abstract Mathematical apparatus and concepts into day to day language. In Classical Physics one knows what is meant by the position or the momentum of a particle and those concepts can be observed and understood without an intermediate interpretation. However, in Quantum Physics, the state of a particle or a system is represented by a vector in Hilbert space and observables are represented by Hermitian operators in Hilbert space. An interpretation or interpretations are needed to express these and other concepts to build a concrete picture out of the abstract apparatus. Copenhagen interpretation is one such interpretation and it is the standard interpretation as far as most of the Physicists are concerned. 
\vspace*{5mm}

Bohr more than anybody else was instrumental in formulating the Copenhagen interpretation, and he in turn was influenced by positivism and Chinese Ying - Yang Philosophy. As a positivist he believed that only the sensory perceptible phenomena exist and did not believe in the existence of that could not be ``observed''. When a state of a particle or system is represented by an eigenvector of an observable (Hermitian operator in Hilbert space) the corresponding value of the observable can be measured and the positivist school had no problem in accepting the existence of such state. For example if the momentum of a particle is known then the state of the particle is represented by a certain vector in Hilbert space, belonging to the particular eigenvalue that has been measured. However, the problem arises when the conjugate Hermitian operator, in this case the position, is considered, as in positivism the ontology is connected with observations and sensory perceptions. We are not considering logical positivism and there seems to be no interpretation of Quantum Physics in a logical positivist ontology.     
\vspace*{5mm}

As we have seen a given eigenstate of a Hermitian operator that has been observed can be expressed as a linear combination of the eigenstates of the conjugate operator. To a positivist, though the given eigenstate exists as it is observed, the eigenstates of the conjugate operator are not observable and it is meaningless for him to talk of such states. Thus if the momentum of a particle has been measured, the eigenstates belonging to the eigenvalues of the conjugate operator, which is the position, are not observed and the positivist would not say anything regarding the existence of such states. As far as the positivist is concerned, there is only a probability of finding the particle at some position, and the particle will be at some position only after the relevant measurement is carried out.  
\vspace*{5mm}

In the case of the double-slit experiment, this means that a positivist would not say whether the particle passes through a particular slit as it is not observed. However he assumes that it it is at one of the slits and not at both as the Aristotelian - Newtonian - Einsteinian ontology demands that the particle should be at one of the slits and not at both slits. (The positivists share with the realists the Aristotelian - Newtonian - Einsteinian ontology. They differ from the realists when they insist that nothing could be said of non observables.)  If a measurement is made, that is if an experiment is carried out to find out the slit where the particle is, then the particle would be found at one of the slits washing out the ``interference pattern''. Then superposition is collapsed and ``decoherance'' sets in resulting ``chaotic pattern''.  
\vspace*{5mm}

A realist differs from a positivist in that the former would want to know the slit at which the particle is (the slit through which the ``particle passes'') even without observing it. He would say the particle would pass through one of the slits whether one observes it or not, and that it is an integral property of the particle independent of the observer. The Classical Physicists were realists. An object in Classical Physics has a momentum whether it is measured or not. The observer in Classical Physics measures the momentum that the particle already possesses. In Quantum Physics the positivists would say that the particle has no momentum before it is measured but acquires a momentum as a result of the measurement. 
\vspace*{5mm}

We would not go into further details on the differences between the realist position and the positivist position. However, what is relevant to us is that both the realist and the positivist would agree that the particle ``goes through one slit'', meaning that at a ``given time'' the particle is found only at one of the slits. They would also agree on the wave nature of the particles. They have to depend on the wave nature as they assume that the particle passes through only one slit, and as such they would not be able to explain the ``interference patterns'' without the wave properties of the particles, as particles ``passing through'' only one slit would not produce ``interference patterns''. 

\section{A NEW INTERPRETATION}
We differ from the positivists as well as the realists since we believe that the particle is found at both slits and hence ``pass through both'' in the common parlance. In general we include the postulate that the eigenstates $|\Psi>$'s in $|\Phi>=\sum|c_i\Psi_i>$  exist in addition to $|\Phi>$ (Postulate 3 below). We have also introduced the concept of a mode. A mode of a particle or a system is essentially a potential observable. A mode has the potential to be observed though it may not be observed at a particular instant.  For example, position, momentum, spin are modes. A particle or a system can be in both modes corresponding to two conjugate Hermitian operators, though only one mode may be observed. 
\vspace*{5mm}

A revised version of the postulates of the new interpretation formulated by Chandana and de Silva$^9$  is given below.  

\begin{enumerate}
\item[1. ] A state of a Quantum Mechanical system is represented by a vector (ray) $\chi$  in the Hilbert space, where $\chi$  can be expressed as different linear combinations of the eigenvectors in the Hilbert space, of Hermitian operators, any operator corresponding to a mode. In other words a state of a Quantum Mechanical system can be represented by different linear combinations of eigenvectors of different modes, each linear combination being that of the eigenvectors of one of the modes. Thus a state could have a number of modes, each mode being a potential observable.
\item[2. ] If $\chi$ is expressed as a linear combination of two or more eigenvectors of a Hermitian operator, that is a mode, then the corresponding mode cannot be observed (or measured) by a human observer with or without the aid of an apparatus. In other words the particular mode cannot be observed and a value cannot be given to the observable, which also means that no measurement has been made on the observable. 
\item[3. ] However, the non observation of a mode does not mean that the mode does not ``exist''. We make a distinction between the ``existence'' of a mode, and the observation of a mode with or without the aid of an apparatus. A mode corresponding to a given Hermitian operator could ``exist'' without being observed. The knowledge of the ``existence'' of a mode is independent of its observation or measurement. In other words the knowledge of the ``existence'' of a mode of a Quantum Mechanical state is different from the knowledge of the value that the observable corresponding to the relevant Hermitian operator would take.
\item[4. ] If a mode of a Quantum Mechanical state is represented by a single eigenvector, and not by a linear combination of two or more eigenvectors, of a Hermitian operator, then the mode could be observed by a human observer with or without the aid of an apparatus, and the value of the corresponding observable (or the measured value) is given by the eigenvalue which the eigenvector belongs to. It has to be emphasised that only those modes of a Quantum Mechanical state, each represented by a single eigenvector, and not by a linear combination of eigenvectors, of an Hermitian operator can be observed at a given instant. 
\item[5. ]If a mode of a Quantum Mechanical state is represented by an eigenvector of a Hermitian operator then the mode corresponding to the conjugate operator cannot be represented by an eigenvector of the conjugate Hermitian operator. It can be expressed as a linear combination of two or more of the eigenvectors of the conjugate operator. This means that the mode corresponding to the conjugate operator cannot be observed, or in other words it cannot be measured. However, the relevant mode ``exists'' though it cannot be observed.
\item[6. ]It is not necessary that at least one of the modes corresponding to two conjugate operators should be represented by a single eigenvector of the relevant operator. It is possible that each mode is represented by linear combinations of two or more eigenvectors of the corresponding operator. In such situations neither of the modes could be observed. 
\item[7. ]A state of a Quantum Mechanical system can be altered by making an operation that changes a mode or modes of the state. However, not all operations correspond to measurements or observations. Only those operations that would result in a mode being expressed as a single eigenvector, and not as a linear combination of the eigenvectors of an operator would result in measurements.  
\item[8. ]A particle entangled with one or more other particles is in general represented by a linear combination of eigenvectors of an Hermitian operator with respect to a mode, while the whole system of particles is in general represented by a linear combination of the Cartesian products of the eigenvectors. In the case of two particles it takes the form  $\sum c_{ij} |\phi_i>|\phi_j>$. If one of the particles is in a mode that is observed, then the particles entangled with it are also in the same mode as an observable. If a measurement is made on some other mode then instantaneously, the corresponding values in the same mode of the entangled particles are also determined. In such case, for two particles the whole system is represented by vectors of the form $|\phi_i>|\phi_j>$. If the number of entangled particles is less than the dimension of the space of the eigenvectors of the Hermitian operator, then if a measurement is made in the particular mode, the particle would be represented by one of the eigenvectors, while the other particles entangled with it would be each represented by a different eigenvector of the Hermitian operator. However, if the number of entangled particles is greater than the dimension of the space of the eigenvectors, then in some cases, more than one particle would be represented by a given eigenvector. 
\end{enumerate}

According to this interpretation if the momentum of a particle is known then it has not one position but several positions. In other words the particle can be at number of positions in superposition though we are not able to observe it at any one of those positions. The particle could be observed only if it is at one position. If an experiment is carried out to determine the position of the particle the superposition or the wave function would collapse, and the particle would be located at one of the positions where it was before the measurement was made. 
\vspace*{5mm}

Similarly if the particle is in the position mode that is observed then it can have several momenta in superposition but we would not be able to observe any one of them. If we perform an experiment to determine the momentum, that is if a measurement is made, then the superposition of momenta would collapse to one of them, enabling us to determine the value of the momentum. 
\vspace*{5mm}

With respect to the double-slit experiment this implies that the particle is at both slits in superposition without being observed and if we perform an experiment to determine the slit ``through which the particle passes'' (the slit where the particle is) then the superposition collapses and the particle would be found only at one of the positions. The positivists while assuming that the particle ``passes through only one slit'' would not say anything on the slit ``through which the particle passes'' as it cannot be observed. For the positivist it is meaningless to speculate on something that cannot be observed. The realists too assume that the particle ``passes through'' only one slit but would not be satisfied with the positivist position, and claim that a theory that is not able to determine the slit through which the particle passes is incomplete.
\vspace*{5mm}

We make a distinction between being in existence and being observed. A particle or a system can exist in a certain mode without being observed. In this case the state of the particle or the state is expressed as a linear combination or superposition of the eigenstates of the relevant Hermitian operator and the particle or the system exists in all the relevant eigenstates without being observed. The mode is observed only when the state of the particle or the system is expressed as a single eigenstate of the relevant Hermitian operator.   
\vspace*{5mm}

The existence of modes with more than one eigenstates has been known for sometime. Monroe$^{10}$  and his colleagues in 1996 were able to demonstrate the existence of two spin states of Beryllium cation simultaneously however without observing them. One could say that the interference obtained by them could be understood on the basis of the existence of simultaneous spin states of the Beryllium cation. Since then similar experiments have been carried out and the existence of superposition of eigenstates cannot be ruled out anymore.

\section{A DIFFERENT ONTOLOGY AND LOGIC}
In the ontology presented here no distinction is made of the existence of sensory perceptible objects and of other entities. There is no absolute existence as such and all existences are relative to the mind. It has been shown by de Silva$^{11}$  that even the mind could be considered as a creation of the mind a phenomenon not in contradiction with cyclic thinking. It is the mind that creates concepts including that of self, and as such sensory perceptible objects do not have any preference over the others.
\vspace*{5mm}

As we have mentioned the positivists find it difficult to take cognizance of entities that are not sensory perceptible and it is this ontology that makes them not to commit on the existence of unobserved ``objects''. In the present ontology all existences are only conventional and not absolute as such. Thus the existence of simultaneous eigenstates or superposition of eigenstates is not ruled out in the present ontology. We have no inhibition to postulate the existence of such states and it is not in contradiction with {\it catuskoti} or fourfold logic that may be identified as the logic of the ontology presented here.  
\vspace*{5mm}

As Jayatilleke$^{12}$  has shown in fourfold logic or sometimes referred to as tetra lemma, unlike in twofold logic a proposition and its negation can be both true and false. (However, we do not agree with the interpretation of fourfold logic given by Jayatilleke.). In twofold logic if a proposition is true then its negation is false, and if a proposition is false, then its negation is true. In addition to these two cases fourfold logic has two more cases where both the proposition and its negation can be true or both false. Thus the proposition that a particle is at $A$, and the proposition that a particle is not at $A$, can be both true in fourfold logic. (According to fourfold logic the case could arise where the particle may be neither at $A$ nor not at $A$.)   We may deduce from that a particle can be both at $A$ and $B$ (not at $A$) at the ``same time''. In other words a particle can be at both slits in respect of the double-slit experiment, and in general a mode represented by a superposition of two or more eigenvectors can exist as the particle or the system can be at number of ``positions'' simultaneously in fourfold logic.
\vspace*{5mm}

In twofold Aristotelian logic a particle has to be either at $A$ or not at $A$.  Thus the Physicists whether they are realists or positivists find it difficult to accept that a particle can ``pass through both slits'' simultaneously, and they have to resort to so called wave nature in order to explain the interference patterns. 

\section{DISCUSSION}
It is seen that both wave picture and the ordinary particle picture fail to explain the interference patterns observed in the double-slit experiment. The wave picture fails as a weak intensity stream of electrons (one electron at a time) exhibits no interference patterns in the case of few electrons. The ordinary particle picture fails as a particle passing through only one slit would not produce interference patterns. The Physicists had to resort to the wave picture as the logic in either positivism or realism would not permit a particle to pass through both slits. 
\vspace*{5mm}

In the case of the experiments conducted by Chandana then at the University of Kelaniya, Sri Lanka, the wave picture as well as the classical particle picture come across more problems as neither a wave nor an ordinary particle would be able to penetrate the aluminium sheets without being affected. These experiments justify our new interpretation involving modes of the particle or the system and the particle picture presented here where a particle can be at both slits. In general we postulate that a particle or system can exist in a mode where more than one eigenstates are in a superposition. The position where a particle is found depends only on the relevant probability, and the so-called interference patterns are only collections of images formed by such particles striking the screen at different positions with the relevant probabilities.  
\vspace*{5mm}

The new postulates are consistent with the ontology where the ``existence'' of a particle or an object does not necessarily mean that it could be observed or that it is sensory perceptible in general, and the fourfold logic. It appears that, unlike Classical Physics with its twofold logic and realist ontology, Quantum Physics is rooted not even in a ``positivist ontology'' but in a different ontology and fourfold logic and we should be able to develop new concepts in Quantum Physics, especially regarding the motion of a Quantum particle from a point $A$ to another point $B$. It is not known how a particle ``moves'' from the double-slit to the screen in the experiments carried out by Chandana, nor how a particle with less energy than the value of a potential barrier ``scales the walls''. In the latter case all that the Physicists have done is to come up with concepts such as ``tunnel effect''. It may be that it is neither the particle that left the point $A$ nor some other particle that reaches the point $B$, if we are to make use of the fourth case of fourfold logic. Chandana in his M.Phil. thesis submitted to the University Kelaniya in September 2008 has described few more experiments that agree with the present ontology and fourfold logic.                  
\vspace*{5mm}
\small
\begin{center}References\end{center}
\hrule
\begin{enumerate}
\item[1.]  Bagget Jim, 1997. The Meaning of Quantum Theory, Oxford University Press.  
\item[2.] Afshar, S.S., 2005. Sharp complementary wave and particle behaviours in the same welcherweg experiment, Proc. SPIE 5866, 229-244. 
\item[3.]Bagget Jim, 1997. The Meaning of Quantum Theory, Oxford University Press.  
\item[4.]Afshar, S.S., 2005.  Sharp complementary wave and particle behaviours in the same welcherweg experiment, Proc. SPIE 5866, 229-244. 
\item[5.]Afshar, S.S., 2005. Sharp complementary wave and particle behaviours in the same welcherweg experiment, Proc. SPIE 5866, 229-244.
\item[6.]Chandana S. and de Silva Nalin, 2004. On the double-slit experiment, Annual Research Symposium, University of Kelaniya, 57-58. 
\item[7.]Chandana S. and de Silva Nalin, 2007. Some experiments involving double-slits, Annual Research Symposium, University of Kelaniya,133-134. 
\item[8.]Bohm D, 1980. Wholeness and the implicate order, Routledge, London. 
\item[9.]Chandana S. and de Silva Nalin, 2004. A new interpretation of Quantum Mechanics, Annual Research Symposium, University of Kelaniya, 59-60. 
\item[10.]Monroe C., Meekhof D. M., King B. E., Wineland D. J., 1996. A ``Schr\"{o}dinger Cat'' Superposition State of an Atom, Science, 272, 1132.
\item[11.]de Silva Nalin, Sinhala Bauddha Manasa  {\it www.kalaya.org/files/nps/070405.pdf}. 
\item[12.]Jayatilleke, K. N.,1963. Early Buddhist Theory of Knowledge, Motilal Banarsidass.

\end{enumerate}

\end{document}